# Two-In-One: A Design Space for Mapping Unimanual Input into Bimanual Interactions in VR for Users with Limited Movement

Two-In-One

Momona Yamagami*

Microsoft Research, Redmond, Washington, USA, my13@uw.edu

Sasa Junuzovic

Microsoft Research, Redmond, Washington, USA, Sasa.Junuzovic@microsoft.com

Mar Gonzalez-Franco

Microsoft Research, Redmond, Washington, USA, margon@microsoft.com

Eyal Ofek

Microsoft Research, Redmond, Washington, USA, eyalofek@microsoft.com

Edward Cutrell

Microsoft Research, Redmond, Washington, USA, cutrell@microsoft.com

John R. Porter

Microsoft, Redmond, Washington, USA, John.Porter@microsoft.com

Andrew D. Wilson

Microsoft Research, Redmond, Washington, USA, awilson@microsoft.com

Martez E. Mott

Microsoft Research, Redmond, Washington, USA, Martez.Mott@microsoft.com

Virtual Reality (VR) applications often require users to perform actions with two hands when performing tasks and interacting with objects in virtual environments. Although bimanual interactions in VR can resemble real-world interactions—thus increasing realism and improving immersion—they can also pose significant accessibility challenges to people with limited mobility, such as for people who have full use of only one hand. An opportunity exists to create accessible techniques that take advantage of users' abilities, but designers currently lack structured tools to consider alternative approaches. To begin filling this gap, we propose Two-in-One, a design space that facilitates the creation of accessible methods for bimanual interactions in VR from unimanual input. Our design space comprises two dimensions, bimanual interactions and computer assistance, and we provide a detailed examination of issues to consider when creating new unimanual input techniques that map to bimanual interactions in VR. We used our design space to create three interaction techniques that we subsequently

---

* Currently at the Department of Electrical & Computer Engineering, University of Washington, Seattle, WA

implemented for a subset of bimanual interactions and received user feedback through a video elicitation study with 17 people with limited mobility. Our findings explore complex tradeoffs associated with autonomy and agency and highlight the need for additional settings and methods to make VR accessible to people with limited mobility.

**CCS CONCEPTS •Human-centered computing~Accessibility~Accessibility theory, concepts and paradigms** •Human-centered computing~Human computer interaction (HCI)~Interaction paradigms~Virtual reality

**ACM Reference Format:**
NOTE: This block will be automatically generated when manuscripts are processed after acceptance

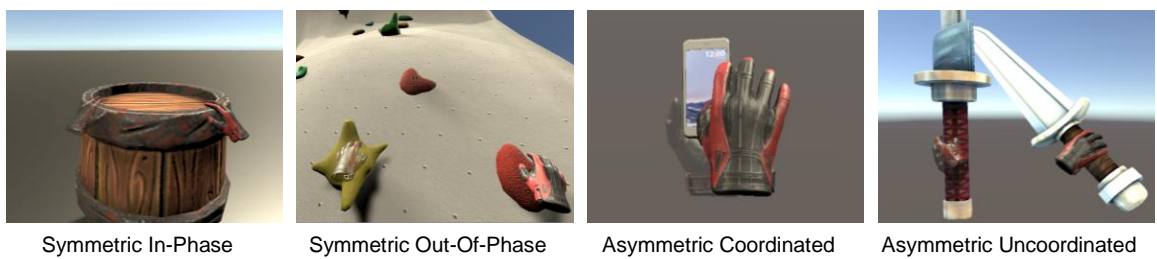

Symmetric In-Phase     Symmetric Out-Of-Phase     Asymmetric Coordinated     Asymmetric Uncoordinated

Figure 1: Bimanual interactions in virtual reality can be categorized into one of four categories (from left to right): symmetric in-phase (e.g., grabbing a heavy object), symmetric out-of-phase (e.g., climbing a wall), asymmetric coordinated (e.g., using a smartphone) and asymmetric uncoordinated (e.g., holding a sword in each hand). Novel input techniques enable the use of one motion controller to perform bimanual interactions.

## 1 INTRODUCTION

Virtual Reality (VR) provides compelling opportunities for immersive digital experiences in numerous domains such as education, gaming, and healthcare. These immersive experiences are delivered through head-mounted displays (HMDs) that occlude users' vision of the physical world and present an assortment of interactive virtual environments (VEs). Users typically interact with objects that populate VEs with hand-held controllers or with bare hands through the use hand-tracking cameras embedded inside HMDs.

Interactions inside VR applications often mirror their real-world equivalents. For example, throwing a virtual bowling ball requires a user to swing their arm backward then forward before releasing it. This style of interaction is in sharp contrast to desktop, mobile, and console gaming systems, where users employ an array of swipes, taps, button presses, and mouse movements to control virtual content. The degree of bodily involvement [Gerling et al., 2021] associated with VR provides a sense of immersion [Ahuja et al., 2021] not found in traditional computing platforms that rely on 2D input.

Interactions in VR, as with interactions in all interactive technologies, presume users possess certain abilities, making it difficult or impossible for users whose abilities do not match the ability-assumptions [Wobbrock et al., 2011] embedded in the design of VR devices and applications to fully use the technology. Recently, researchers have investigated the accessibility—or lack thereof—of VR for people with physical disabilities and found significant accessibility barriers that can prevent participation [Gerling et al., 2020; Gerling and Spiel, 2021; Mott, 2020]. Some barriers, such as inaccessible buttons on VR controllers [Gerling et al. 2020; Mott, 2020], are common nuisances regularly encountered by people with physical disabilities, while other barriers are endemic to the full-body gestural nature of VR interaction. Specifically, simultaneously manipulating two hand-held controllers can pose significant challenges to people who may have had an amputation of an arm, a stroke limiting movement in one hemisphere of their body, or to individuals who might have a hand preoccupied with controlling their power wheelchair. VR applications often include experiences where users control a two-handed avatar and perform bimanual interactions, and as a result, these applications are rendered inaccessible to people who, by either choice or necessity, prefer to use a single VR controller.

Allowing users with limited mobility to perform bimanual interactions in VR with a single controller could significantly improve the accessibility of VR applications. However, it is unclear how designers and developers should enable these experiences and what tradeoffs users with limited mobility might experience when engaging with VR applications. Creators of VR experiences need structured methods for developing accessible alternatives to bimanual interactions.

To help facilitate the creation of these alternatives, we propose Two-in-One, a design space for mapping unimanual input into bimanual interactions in VR. Our design space consists of two dimensions: (1) bimanual interactions (Figure 1; symmetric in-phase, symmetric out-of-phase, asymmetric coordinated, asymmetric uncoordinated) and (2) computer assistance (on, off). We first describe four types of bimanual interactions based on Guiard's Kinematic Chain theory [Guiard, 1987] and classify bimanual interactions found in existing VR games and applications into the four categories. We then introduce two different views on the design space (creation and evaluation lens) that can be used to fill in the contents of the cells created by the two dimensions. We apply the two different lenses on the design space to create interaction techniques that enable more accessible control of bimanual interactions in VR. Lastly, we present the results of a remote video elicitation study with 17 participants with limited mobility who provided feedback on three prototype interaction techniques we implemented.

Our work makes the following:
1. a design space, Two-in-One, for mapping unimanual input into bimanual interactions in VR, offering opportunities to devise accessible interaction techniques that take advantage of users' abilities;
2. development and demonstration of using the creation lens to identify three interaction techniques with the potential to enable accessible control of bimanual interactions;
3. development and demonstration of applying the evaluation lens to predict tradeoffs between the different input techniques and verifying the predictions based on feedback from 17 people with limited mobility for one bimanual interaction (symmetric out-of-phase) on their thoughts, preferences, and tradeoffs associated with three prototype interaction techniques.

Two-in-One provides designers a tool for developing interaction methods for VR games and applications that are accessible to people with limited mobility, and our user study highlights the need for a broad and inclusive set of accessibility settings for personalizable immersive experiences in VR.

## 2 RELATED WORK

Our research was motivated by investigations of bimanual actions in daily life and during computer use, previously identified accessibility challenges with VR technology for people with limited mobility, and the limited number of accessibility options currently available in VR. We leverage ideas from prior work in design spaces for human-computer interaction (HCI) to derive our design space.

## 2.1 Bimanual Interactions in Daily Life

Accelerometry data suggests that individuals without limited mobility use their dominant and non-dominant hands approximately equally and that most interactions in daily life require the simultaneous use of both hands [Lang et al., 2017]. Standardized handedness questionnaires that ask people how they perform common tasks of daily living place heavy emphasis on bimanual activities, such as striking a match or using a broom [Annett, 1970; Oldfield, 1971], and studies on recovery after stroke suggest that regaining the use of both hands is crucial for performing activities of living independently [Haaland et al., 2012; Michielsen et al., 2012; Vega-González et al., 2005]. While traditional desktop and mobile computing systems often rely on unimanual input—primarily for pointing, selection, and performing unistroke gestures—it is reasonable to expect interactions in VR to heavily emphasize bimanual input as VR technologies continue to mimic properties of the physical world inside virtual environments. Although bimanual interactions in the real world might require physical adaptations to ensure accessibility for users with limited mobility, virtual environments can be manipulated to enable accessible bimanual interactions with unimanual input that would not be possible in real life.

## 2.2 Bimanual Interactions in HCI

Traditional computer interfaces rely heavily on pointing and clicking using one-handed input (e.g., pointing with a mouse or tapping a touchscreen with a finger). Parallelizing certain computer interactions by using both hands simultaneously has been shown to improve performance while not significantly increasing cognitive load [Buxton et al., 1986; Hinckley et al., 1998; Kabbash et al., 1994; Yee, 2004]. For example, participants spontaneously used both hands in parallel and improved performance during an object matching task when the researchers enabled object tracking with one hand while scaling the size of the object with the other hand [Buxton and Myers, 1986]. However, performance improvements from parallelizing movements are limited to when the hands perform *coordinated tasks*, where one hand serves as a frame of reference and the other hand performs an action [Guiard, 1987]. When participants were asked to perform two separate tasks in parallel, such as manipulating two cursors using two mice, their performance degraded even more than if they had done the same task with just one hand [Kabbash, Buxton and Sellen, 1994]. Recent work on context-aware sensing suggests that handheld devices like smartphones and tablets benefit from adapting to inputs provided by both hands in a coordinated fashion by using input from both the hand holding the device and the hand actively using it [Hinckley et al., 2000; Zhang et al., 2019].

Although one-handed interactions are commonly used in traditional interactions with computers, bimanual interactions are popular in VR since hand movements can be directly mapped from the real to the virtual world. Work in VR on two-handed gestural and manipulation input techniques based on coordinated movement suggests that certain situations and practice allow for more efficient two-handed interaction compared to single-handed interaction [Lévesque et al., 2013; Veit et al., 2008]. Researchers attempted to make gestures and interactions as similar to the real world as possible (e.g., select by performing thumbs up with the non-dominant hand) and demonstrated that when the dominant hand is used to perform precise tasks and the non-dominant hand is used to perform stabilizing movements, people are faster at performing selection tasks with two hands compared to one hand [Lévesque, Laurendeau and Mokhtari, 2013]. As VR technology continues to improve in fidelity and realism, bimanual interactions will be an increasingly popular interaction method and it is important to ensure that such bimanual interactions are accessible to people with different abilities.

## 2.3 Accessibility Challenges in VR for Users with Limited Movement

Researchers have found that people with limited mobility experience several accessibility challenges when using VR. Mott et al. identified seven accessibility barriers related to the physical accessibility of VR devices, which included barriers such as putting on and taking off HMDs and difficulty manipulating two motion controllers [Mott, 2020]. In their critical examination of VR technology, Gerling and Spiel described how the level of "bodily involvement" places a heavy emphasis on the movements of various body parts (e.g., head, arms, hands, torso, etc.), which can present barriers to people with physical disabilities [Gerling and Spiel, 2021]. Gerling and Spiel also described how the actions lent during interactions with VR systems can pose numerous access barriers, such as fatigue from extended use [Gerling and Spiel, 2021]. Franz et al. addressed one aspect of bodily involvement—the requirement that users rotate their heads or bodies to view a virtual scene—by developing Nearmi, a framework for designing techniques that allow people with limited mobility to identify and orient to points of interests in virtual environments [Franz et al., 2021].

Manipulating two motion controllers can be especially challenging for people with limited mobility [Mott et al., 2019; W3C, 2020]. Some people may only use one hand to hold a controller due to stroke or amputation, and other people may use one hand so they can drive their wheelchair with the other hand [Mott, 2020]. One approach to making dual controller interactions more accessible is to add a level of automation to them. However, complete automation of character movements that are difficult for users to perform is also not desirable. When participants tested VR games with automated movements designed for players who use wheelchairs, some felt that the experience lacked embodiment and preferred to control their avatar's movements themselves [Gerling et. al, 2020]. Thus, enabling individualized adaptive accessibility tools is an important aspect of ensuring accessibility in VR for users with limited mobility [Gerling et. al, 2020; Mott et. al, 2019; Mott, 2020].

## 2.4 Current Options for Enabling Accessible Use of VR Technologies

VR games remain difficult to access out of the box for people with limited mobility. Grassroots efforts such as WalkinVR [2MW, 2002], Able Gamers [charity, 2020], and Special Effect [2020] provide users with software, tools, and equipment to adapt gaming equipment and applications to match their abilities. However, it is often up to developers to ensure that games and other applications are accessible to people with disabilities. For example, VR accessibility guidelines from W3C [W3C, 2020], as well as more general game accessibility guidelines ["Game Accessibility Guidelines", 2020], are useful for creating games that have motor, visual, and hearing accessibility options. Console games such as "The Last of Us Part II" [LLC, 2020] and VR games such as "Half Life Alyx" [Valve, 2020] demonstrate the possibilities for embedding accessible features directly into games for a wide range of abilities.

Enabling bimanual interactions in VR with a single motion controller would improve VR accessibility for people with limited mobility by allowing them to personalize their gaming and interaction experience. Regardless of whether this is done at a first- or a third-party level, what is needed to activate such accessible experiences is a design space that can provide guidance in mapping unimanual input into bimanual interactions in VR.

## 2.5 Design Spaces in HCI

Design spaces and taxonomies have been used within HCI to categorize existing technologies [Card et al., 1990; Mackinlay et al., 1990], to synthesize knowledge across domains [Brudy et al., 2019], and to develop novel interaction methods and input techniques [Hirzle et al., 2019; Pfeuffer et al., 2014]. Mackinlay [Mackinlay et. al, 1990] and Card [Card et. al, 1990] constructed design spaces to categorize input devices. By doing so, they identified essential properties that applied to existing and future input devices. In their subsequent work, Card et al. [Card et al., 1991] performed a morphological analysis to generate a new design space for input devices, and to identify points within the space that warranted further investigation. Brudy et al. [Brudy et. al, 2019] proposed a cross-device taxonomy that synthesized related work from multiple related domains within HCI to inform future research in this area. Pfeuffer et al. [Pfeuffer et. al, 2014] constructed a design space for direct-touch and gaze-touch interaction techniques to understand similarities and differences between the two approaches. In their exploration of gaze interactions on head-mounted displays, Hirzle et al. [Hirzle et. al, 2019] used three views of their design space to demonstrate different ways cells within the space could be filled. Views demonstrate the versatility of design spaces by providing a *lens* for researchers to examine different technological or social aspects of the space. In this work, we develop and use two lenses – the creation and evaluation lens – to demonstrate how our Two-In-One design space can be used to *create* and *evaluate* accessible input in VR.

Two-In-One follows in the tradition of these works—and others—by providing a structured way to develop methods that afford more accessible bimanual interactions in VR.

## 3 TWO-IN-ONE: DESIGN SPACE FOR MAPPING UNIMANUAL INPUT INTO BIMANUAL INTERACTIONS IN VR

Our Two-In-One design space is a tool for creating and evaluating different input techniques that can map unimanual input into bimanual interactions. We assume that the user is controlling one hand in VR with a motion controller (the *controlled hand*) and the other hand in VR is controlled by an input technique (the *virtual hand*). The two dimensions that span the design space are (1) bimanual interaction types and (2) computer assistance (Table 1). Bimanual interaction types enable VR application developers to place all VR bimanual interactions into one of four interaction types. This allows developers to leverage different properties of each interaction type during input technique creation. Computer assistance enables VR application developers to consider how properties of each interaction type can take advantage of advanced computing methods to offload user effort to the computer and to consider the tradeoffs associated with the input techniques during evaluation.

Table 1: The design space highlight shared properties of a group of bimanual interactions.

| Bimanual interaction / Computer assistance | Symmetric In-Phase (e.g., move heavy object) | Symmetric Out-Of-Phase (e.g., climb ladder) | Asymmetric Coordinated (e.g., use smartphone) | Asymmetric Uncoordinated (e.g., use two swords) |
|---|---|---|---|---|
| Computer Assistance On | | | | |

| Computer Assistance Off | | | | |

## 3.1 Dimension 1: Bimanual Interactions in VR

The first dimension of our design space includes the types of bimanual interactions that occur in VR. Consider a VR application developer who wants to make their application more accessible by supporting one-handed input. Without this dimension, the developer must identify all possible bimanual interactions and individually craft a one-handed solution for each interaction. With our design space, however, the developer can categorize the VR bimanual interactions into one of four categories and develop one-handed solutions for each interaction type. We adapt Guiard's Kinematic Chain theory on real-world bimanual interactions [Guiard, 1987; Hinckley et al., 1997; Kabbash et. al, 1994] to propose that all VR bimanual interactions can be categorized into one of four categories: *symmetric in-phase, symmetric out-of-phase, asymmetric coordinated,* and *asymmetric uncoordinated*.

Guiard's Kinematic Chain theory on real-world bimanual interactions suggests that interactions are either symmetric, where the two hands perform the same movement, or asymmetric, where the two hands perform separate movements [Guiard, 1987; Hinckley et. al, 1997; Kabbash et. al, 1994]. Symmetric interactions can be further broken down by whether the two hands perform the same movement at the same time (*in-phase)* or at different times *(out-of-phase)*. Activities such as lifting a heavy object with two hands are considered symmetric in-phase interactions, whereas climbing a ladder or rope is considered symmetric out-of-phase interactions.

We adapt Guiard's theory by separating asymmetric interactions into *coordinated interactions* [Guiard, 1987] and *uncoordinated interactions*. For asymmetric coordinated interactions, one hand (typically the non-dominant hand) provides a stabilizing frame of reference, while the other hand (typically the dominant hand) performs some movement [Guiard, 1987]. For example, when holding a smartphone with one hand and typing with the other, the hand holding the smartphone acts as the stabilizing hand while the other hand performs a movement by typing on the device. However, in asymmetric uncoordinated interactions, the movement of one hand is not dependent on the movement of the other. For example, when using two swords in a game such as Beat Saber, the slashing movement of one hand is not dependent on the movement of the other hand. In Guiard's Kinematic Chain theory, asymmetric uncoordinated interactions are considered two unrelated unimanual interactions performed by two hands, but to keep within our bimanual design space, we redefined these interactions as bimanual asymmetric uncoordinated interactions.

In conclusion, we hypothesize that all VR bimanual interactions can be categorized into one of four interaction types:

1) *Symmetric in-phase*: The left and right hand perform the same or similar movement synchronously (e.g., jump-roping, turning a wheel).
2) *Symmetric out-of-phase:* The left and right hand perform the same or similar movements one after the other in series, or asynchronously (e.g., climbing a ladder, pulling a rope).
3) *Asymmetric coordinated:* The left and right hands perform separate movements on related objects where the non-dominant hand acts as a stabilizing hand and the dominant hand performs an action (e.g., striking a match, swinging a golf club).

4) *Asymmetric uncoordinated:* The left and right hand perform separate actions on two unrelated objects where each hand moves independently from one another and the movement of one hand has no relation to the movement of the other hand (e.g., using two swords at the same time, opening a door while using a phone).

*3.1.1 Survey of Bimanual Interactions in Popular VR Applications*

We surveyed popular VR applications to understand common VR bimanual interactions more deeply and to determine whether VR bimanual interactions can be categorized into the four categories we proposed. We surveyed the top 25 most popular VR applications on August 7, 2020, that required handheld motion controllers in the Steam and Oculus stores (39 VR applications in total due to overlap between the two stores) and watched at least 30 minutes of application use on YouTube by at least two players (see supplementary materials for more information).

We observed at least one bimanual interaction in all but one application (Virtual Desktop). All observed bimanual interactions were successfully categorized into one of the four categories, and no bimanual interaction was placed in more than one category. Asymmetric coordinated interactions were about half of all observed bimanual interactions, with symmetric in-phase interactions being the least observed (Table 2). As many of the popular VR applications were shooting or fighting games, many of the interactions involved using a weapon (e.g., asymmetric coordinated: loading a gun, asymmetric uncoordinated: using two weapons at the same time). Climbing comprised almost 75% of all symmetric out-of-phase interactions and grabbing a ledge with both hands and moving heavy objects comprised over 50% of all symmetric in-phase interactions. The survey demonstrates that the four bimanual interaction categories described in the first dimension of our design space convey essential properties about each group of interactions, which made it possible to categorize all observed VR bimanual interactions into distinct categories.

Table 2: Most common bimanual interactions identified in popular VR games.

| **Symmetric In-Phase** *(14%; 20 occurrences)* | **Symmetric Out-Of-Phase** *(19%; 26 occurrences)* | **Asymmetric Coordinated** *(48%; 66 occurrences)* | **Asymmetric Uncoordinated** *(19%; 26 occurrences)* |
|---|---|---|---|
| Jump and grab ledge with both hands *(35%; 7 occurrences)* | Climb object *(73%; 19 occurrences)* | Other *(33%; 22 occurrences)* | Use two weapons at the same time *(77%; 20 occurrences)* |
| Move heavy object with both hands *(35%; 7 occurrences)* | Swings arms to move around *(15%; 4 occurrences)* | Load gun *(18%; 12 occurrences)* | Perform action while holding object with other hand *(19%; 4 occurrences)* |
| Movement *(15%; 3 occurrences)* | Pull rope or chain *(11%; 3 occurrences)* | Hold weapon with two hands *(18%; 12 occurrences)* | Other *(4%; 2 occurrences)* |
| Other *(15%; 3 occurrences)* | | Do something to wrist with other hand *(8%; 5 occurrences)* | |
| | | Navigate phone or tablet *(8%; 5 occurrences)* | |

| | |
|---|---|
| | Using a bow and arrow *(8%; 5 occurrences)* |
| | Pull pin out of grenade *(8%; 5 occurrences)* |

### 3.2 Dimension 2: Computer Assistance

The second dimension of our design space considers how the actions of the virtual hand should be controlled by an input technique. Due to the digital nature of VR, opportunities exist for VR application developers to offload user effort via computer assistance. Sub-disciplines within HCI have explored computer-aided approaches for improving users' experiences by decreasing user effort during tasks such as information search and retrieval [Cai et al., 2016], coding [Kim et al., 2021], and VR interaction [David-John et al., 2021]. These approaches, however, have posed problems since their conception because they can be frustrating or detrimental to completing tasks if the computer cannot correctly predict what the user intended [Olteanu et al., 2020; Yang et al., 2020]. Although computer-aided approaches do pose some risk, there are opportunities to leverage them to create innovative solutions.

It is important to understand the tasks and users' goals when developing computer-aided approaches. Horvitz proposed twelve principles for the creation of mixed-initiative user interfaces that leverage both user input—typically through direct manipulation—and automated reasoning [Horvitz, 1999]. These principles provide best practices developers should follow when considering when and how computers should assist users. For example, the principle of *developing significant value-add automation* suggests that if we are adding computer assistance, there should be a positive tradeoff such as decreased user effort or quicker interactions. Additionally, the principle of *allowing efficient direct invocation and termination* suggests that making it possible for the user to turn the computer assistance on and off would be beneficial for assisted interaction [Horvitz, 1999]. With our design space, these considerations will be limited to specific bimanual interactions. As a result, developers can explore tradeoffs for when computer assistance should be on or off depending on the task and the bimanual action being performed.

When *computer assistance is on*, the virtual hand will not be directly controlled by the user but will instead be controlled by computer assistance that predicts where the user wants the virtual hand to be. The benefit of computer assistance is that the user can solely focus on manipulating the controlled hand while the computer takes control of the virtual hand. The logic determining the behavior of the virtual hand can be programmed heuristically or algorithmically with the potential to leverage advances in machine learning [Yang et. al, 2020]. Computer assistance can be helpful to users as they only need to manage their controlled hand, which decreases user effort. However, computer assistance could lead to a feeling of loss of autonomy, as the virtual hand might be placed in undesirable locations or perform the bimanual interaction differently from how the user intended.

When *computer assistance is off*, the user must directly control their virtual hand in some way through an input technique. This additional control could be beneficial because the user can directly interact with the VR environment, thus increasing autonomy but also potentially increasing user effort. In our evaluation lens section, we demonstrate through a user study the extent to which the computer assistance dimension can predict tradeoffs that users may experience when using these approaches.

## 4 CREATION LENS

Two-In-One can be best utilized for creating and evaluating accessible VR interactions by applying different views to the cells within the design space. Following Hirzle et. al. [Hirzle et. al, 2019]'s previous work on developing three different views to utilize their design space, we developed two views – the creation and evaluation lens – to assist researchers and developers in creating and evaluating accessible VR interactions. Each cell is characterized by a bimanual interaction (symmetric in-phase, symmetric out-of-phase, asymmetric coordinated, asymmetric uncoordinated) and computer assistance (on, off) for a total of eight cells. With the creation lens, we highlight properties of each cell that enable development of potential input techniques for groups of interactions. With the evaluation lens, we highlight tradeoffs that helps make a preliminary prediction of the benefits and drawbacks that people may feel when using different prototyped input techniques.

### 4.1 Bimanual Interaction Properties

The creation lens of the design space allows designers to identify and leverage different properties of each bimanual interaction to develop potential input techniques. Each bimanual interaction has two binary properties, coordinated versus uncoordinated action and synchronous versus asynchronous action, that can be leveraged to develop input techniques (Table 3).

Table 3: The design space highlights shared properties of a group of bimanual interactions.

| Bimanual interaction / Computer assistance | Symmetric In-Phase (e.g., move heavy object) | Symmetric Out-Of-Phase (e.g., climb ladder) | Asymmetric Coordinated (e.g., use smartphone) | Asymmetric Uncoordinated (e.g., use two swords) |
|---|---|---|---|---|
| Computer Assistance On | Coordinated actions | | | Uncoordinated actions |
| Computer Assistance Off | Synchronous actions | Asynchronous actions | Synchronous and asynchronous actions | |

*4.1.1 Coordinated versus Uncoordinated Actions*

The coordinated or uncoordinated property identifies whether the virtual hand's movements can be predicted from the movements of the controlled hand. Movement coordination is useful for designing computer assisted input techniques. For example, because both hands perform the same or similar movements in symmetric interactions, input techniques supporting symmetric in-phase and symmetric out-of-phase interactions could take advantage of symmetric coordinated movements. Similarly, input techniques for asymmetric coordinated interactions should consider how the dominant and non-dominant hands support each other when performing asymmetric coordinated tasks. By definition, in asymmetric coordinated interactions the non-dominant hand acts as a stabilizer—typically by holding an object—while the dominant hand is active by performing an action on the object. To provide the greatest autonomy and ease of movement to the user, the active hand should be the controlled hand holding the motion controller and the stabilizing hand should be the virtual hand. Lastly, asymmetric uncoordinated interactions are not coordinated, and thus the input technique must enable independent movement of the controlled and virtual hands. As much as possible, input techniques should consider how the movement coordination property can be incorporated into the technique to enhance realism and support the user's expectations of how the dominant and non-dominant hands should or should not be coordinated with each other.

*4.1.2 Synchronous versus Asynchronous Actions*

The synchronous or asynchronous property identifies whether the two hands are moving at the same time (synchronously) or at different times (asynchronously). Synchronization is useful when determining whether an input technique must enable synchronous movements. For example, symmetric in-phase interactions are characterized by both hands moving at the same time. Thus, an input technique for such interactions must enable synchronous movement. However, asymmetric coordinated and uncoordinated interactions do not necessarily require synchronous movements, so these interactions could be enabled by input techniques where hand movements occur asynchronously or synchronously. Lastly, symmetric out-of-phase interactions are characterized by the two hands moving one after another, so an input technique for such interactions must enable movements asynchronously. If we are trying to mimic movements in real life that are synchronous in VR, input techniques that enable synchronous movement may enhance realism and embodiment, while input techniques that require asynchronous movement may not feel as realistic.

**4.2 Applying Creation Lens to Develop Input Techniques**

We provide an example of how we applied our Two-in-One design space to generate potential input techniques (Table 4). When computer assistance is on, we can take advantage of the coordination property to infer the movement of the virtual hand. For all interactions except asymmetric uncoordinated, we brainstormed that because these movements are coordinated, a heuristic-based inference or a machine learning-based inference can be used for the virtual hand. An example of a heuristic-based inference could be that most people would hold a bow X inches away from their headset. An example of a machine learning-based inference could be to create a dataset of people using a bow and arrow with two hands and develop an algorithm to predict where the stabilizing hand should be based on gaze and headset position. For asymmetric uncoordinated movements, using computer assistance may be challenging because the movements of one hand are not coordinated with the other. Here, we could automate the movement of the virtual hand heuristically or with machine learning. An example of a heuristic-based inference could be to slash vertically with a sword every X seconds. An example of a machine learning-based inference could be to create a dataset of people using two swords and develop an algorithm to predict how people usually use a sword. However, because we cannot fully predict the movement of the virtual hand based on the movement of the controlled hand, there is a risk that the user might find the uncontrolled and unpredictable movements of the virtual hand distracting or interfering with their enjoyment of VR.

When computer assistance is off, the property of synchronous actions limits potential input techniques for interactions where the two hands must move synchronously (symmetric in-phase). For these interactions, we brainstormed that an alternative input such as head or eye gaze could enable synchronous virtual hand control. Another possible input technique we generated was enabling adaptations similar to real life – people with limited mobility make many one-handed adaptations during everyday bimanual interactions that they could also use in VR. This input technique could enable a user-friendly experience with no learning curve but may be challenging to implement because the adaptations would require sophisticated physics engines and novel methods to sense different body parts other than the hands, which are areas for future research. Lastly, copiloting, where another user controls the virtual hand directly is another possible input technique, but not ideal because it would require another person to be playing with the individual. If the bimanual interaction can be completed asynchronously (asymmetric coordinated, symmetric out-of-phase, asymmetric uncoordinated), we brainstormed that this would enable additional input techniques, as the two hands do not need to move together at the same time. We brainstormed that a mode switch, where the user controls one hand then the other asynchronously could be a useful input technique to control the virtual hand directly.

Table 4: Example brainstorm of potential input techniques to enable bimanual interactions with one motion controller. We discuss the three bolded and italicized input techniques in further detail below.

| Bimanual interaction / Computer assistance | Symmetric In-Phase (e.g., move heavy object) | Symmetric Out-Of-Phase (e.g., climb ladder) | Asymmetric Coordinated (e.g., use smartphone) | Asymmetric Uncoordinated (e.g., use two swords) |
|---|---|---|---|---|
| Computer Assistance On | ***Infer virtual hand heuristically*** or with machine learning | ***Infer virtual hand heuristically*** or with machine learning | | Automate virtual hand using machine learning |
| Computer Assistance Off | Alternative input  Adaptations similar to real life  Copilot | ***Alternative input***  Adaptations similar to real life  ***Mode switch***  Copilot | | |

Below, we discuss three input techniques in further detail: 1) inferring the virtual hand from the controlled hand, 2) using alternative input such as head or eye gaze to control the virtual hand, and 3) mode switching from controlling one hand to another to enable bimanual interactions given one motion controller input (Table 4). We picked these three because they represent potential input techniques with computer assistance on or off that take advantage of the properties of synchronicity and coordination. Although the proposed input techniques do not cover all possible approaches that could be conceived to enable bimanual interactions in VR, these three techniques illustrate how the Two-In-One design space can support the development of input techniques that enable bimanual interactions using one motion controller. However, not all input techniques work for all bimanual interactions, and the design space can be applied to each input technique to determine whether the method would work for a certain group of interactions.

*4.2.1 Potential Input Technique 1: Infer Virtual Hand*

The coordination property indicates that the movements of the virtual hand can be directly inferred for symmetric interactions. For symmetric in-phase interactions, rotational symmetry with respect to the object being interacted with (e.g., turning a wheel) or plane symmetry with respect to the center of the body (e.g., grabbing a ledge with both hands) can be leveraged to infer the movement of the virtual hand from the movement of the controlled hand. For symmetric out-of-phase interactions, the movement timing of the virtual hand must additionally be determined heuristically or algorithmically. For instance, in scenarios such as climbing a ladder, rope, or pipe, to make the movements realistic, the direction, speed, and distance of the controlled hand could be mimicked if the holds are equally spaced. If the holds are not evenly spaced, the placement of the next hand may need to be inferred from the location of the closest graspable object. Lastly, the movement of the stabilizing hand for asymmetric coordinated interactions could be inferred heuristically or determined using datasets. For example, if a person was swinging a golf club, it could be inferred that the virtual hand is placed on the golf club below the controlled hand to act as the stabilizing hand.

*4.2.2   Potential Input Technique 2: Alternative Input*

Using an alternative input such as a joystick, headset position, or eye gaze to control the virtual hand is possible for all four types of bimanual interactions because movements can be performed synchronously or asynchronously, and movement coordination can be built in by constraining the possible locations of the virtual hand. To allow users to fully control movements with the alternative input, it should ideally have at least 6 degrees of freedom (DOF) like the motion controller (e.g., using a foot). If the alternative input does not have 6 DOF (e.g., head gaze, eye gaze, joystick), ray casting methods could be used to control the virtual hand. For symmetric in-phase or out-of-phase interactions, ray casting could be useful to visualize the virtual hand onto an object that is already held in place by the controlled hand (e.g., holding onto a heavy object) or onto an immobile object (e.g., a climbing wall). To further take advantage of the coordination property, the possible locations for the virtual hand could be limited to locations that are approximately symmetric to the controlled hand. For asymmetric coordinated interactions, ray casting may be useful to control the stabilizing hand position at a certain distance away from the user to take advantage of the property of movement coordination. For example, if the user was controlling the position of a smartphone or tablet with their headset, the smartphone could be placed within reaching distance of the controlled hand. Lastly, for asymmetric uncoordinated interactions, depending on the type of object the user is interacting with, it may be difficult to fully control the virtual hand with an alternative input that has less DOF than the motion controller. For certain objects that are only being held (e.g., a shield), holding and controlling the object and virtual hand a certain distance away from the user using ray casting methods may be sufficient. However, for objects that require active movement in all 6 DOF like a sword, it may be difficult to fully control due to insufficient degrees of freedom.

*4.2.3   Potential Input Technique 3: Mode Switch*

In the mode switch input technique, users switch back and forth from controlling the two VR hands one after another. As such, the synchronicity property indicates that symmetric in-phase interactions cannot be completed with this input technique. Mode switching is useful for symmetric out-of-phase and asymmetric coordinated and uncoordinated interactions that do not necessarily require synchronous movement. For asymmetric coordinated interactions, the movement coordination principle suggests that the VR hand corresponding to the hand holding the motion controller should be the active hand and the VR hand corresponding to the hand not holding the motion controller should be the stabilizing hand. Additionally, it may be useful for the active hand to move in coordination with the stabilizing hand when the hand holding the motion controller is controlling the stabilizing hand. This is because, for asymmetric coordinated interactions, the active hand usually interacts directly with the object that is being held by the stabilizing hand (e.g., tapping a smartphone screen).

## 5 EVALUATION LENS

Designers can use the evaluation lens of the design space to predict what tradeoffs or challenges may occur for a created input technique (Table 5). Consider a VR application developer or designer who has created different input techniques for a particular bimanual interaction and wants to understand the inherent tradeoffs associated with the different input techniques. Understanding the tradeoffs can help determine which input technique would be better suited for a particular task and to understand what tradeoffs would be preferable for the application being developed. Additionally, the evaluation lens can be used to narrow down the potential input techniques to prototype and test on potential users.

Table 5: Two tradeoffs (user effort and autonomy) that can be considered using the evaluation lens.

| Bimanual interaction / Computer assistance | Symmetric In-Phase (e.g., move heavy object) | Symmetric Out-Of-Phase (e.g., climb ladder) | Asymmetric Coordinated (e.g., use smartphone) | Asymmetric Uncoordinated (e.g., use two swords) |
|---|---|---|---|---|
| Computer Assistance On | Requires less effort from user, decreases user autonomy | | | |
| Computer Assistance Off | Requires more effort from user, increases user autonomy | | | |

### 5.1 Applying Evaluation Lens to Predict Input Technique Tradeoffs

The extent to which the tradeoff affects the user is dependent on the implementation of the input technique. Applying the evaluation lens to our proposed input techniques, we can assess that inferring the virtual hand (computer assistance on) will require less effort from the user but decrease user autonomy, whereas alternative input and mode switch (computer assistance off) will require more effort from the user but increase user autonomy. In a game setting during a symmetric out-of-phase climbing task like climbing a wall for example, a developer can ideate that while less user effort and decreased autonomy may be preferable for a simple task like climbing a featureless wall (Figure 2a), increased user autonomy despite requiring more user effort may be preferable for a more complex task like going up a climbing wall that requires strategy and decision making (Figure 2b). Below, we ideate predicted tradeoffs for each input technique in further detail.

#### 5.1.1 Predicted Tradeoffs: Infer Virtual Hand

A benefit of using computer assistance to control the virtual hand is that it decreases the amount of effort and fatigue experienced by the user since they only control the position of the controlled hand. The movements of the virtual hand can be performed synchronously with the movements of the controlled hand, improving user performance while playing a time-sensitive application or game. For instance, simple interactions like climbing a ladder or lifting a heavy object may be faster and more intuitive to complete with this input technique. However, the user has less control over the movement of the virtual hand which may decrease the feeling of embodiment and autonomy felt by the user. Thus, depending on user preferences, more complex interactions like scaling a climbing wall or grabbing onto a ledge may benefit from providing more control to the user.

*5.1.2 Predicted Tradeoffs: Alternative Input*

The alternative input technique has the benefit of providing the user with a high level of autonomy and control over the virtual hand and enables the user to perform movements synchronously. For example, the user could aim a bow with the headset while pulling back the arrow simultaneously with the controlled hand. However, the alternative input technique may require the same DOF as the motion controller to fully control the virtual hand. Currently, motion controllers typically have at least 6 DOF, but as hand tracking becomes more complex and the DOF increases with finger individuation, it may be difficult to use certain alternative inputs to control the virtual hand. Additionally, alternative input could cause overloading of information or occlusion of important visual information if the user wants to look around without controlling a virtual hand. Moreover, some alternative inputs may not be accessible to all users. For example, head gaze may not be accessible for users with limited neck movement.

*5.1.3 Predicted Tradeoffs: Mode Switch*

The mode switch technique enables full control over the virtual hand. Because the main assumed ability is that the user can control one hand, it may be more accessible to a wider range of abilities compared to alternative input. However, it could be fatiguing to use one hand to control both hands. Moreover, it could increase task completion times, which could negatively impact performance if timely movements are required. To enable users to complete bimanual interactions that are time-sensitive, it may be helpful to incorporate a slow-down of virtual time when mode switch is enabled. Finally, movements cannot be performed synchronously when using mode switch, which may cause issues with embodiment. Mode switch is a useful input technique for applications where timeliness of an interaction is not crucial and most of the interactions are performed with the dominant hand, such as 3D drawing applications.

**5.2 Prototyping Input Techniques**

To demonstrate the extent to which the tradeoffs predicted using the evaluation lens are or are not reflective of actual user feedback, we prototyped and created videos of *inferring virtual hand*, *alternative input*, and *mode switch* techniques proposed above to elicit user responses. To keep our pilot study to a reasonable time frame (~1 hour) and because this study is a proof-of-concept, we solely prototyped and elicited user responses for symmetric out-of-phase interactions. We designed our remote video elicitation study to minimize physical contact with potential participants due to COVID-19 restrictions affecting in-person user studies. Our results suggest that although the evaluation lens is valuable for getting a general idea of the tradeoffs associated with each input technique and to help narrow down options, a user study is necessary to ensure that the input techniques generated are truly accessible to people with limited mobility.

We implemented prototypes of the three input techniques for two instances of symmetric out-of-phase interactions that we evaluated with people with limited mobility to learn about user preferences for these tradeoffs. One of the interaction instances was a featureless wall where the user could choose anywhere on the wall to climb, and the other instance was a climbing wall with variably spaced holds where the user can only grasp onto the holds to climb (Figure 2). We prototyped two instances of climbing to investigate whether user preferences would change depending on a particular instance, even if the interaction is similar. A more in-depth and in-person exploration of the other interactions is required in the future to fully characterize user needs and preferences.

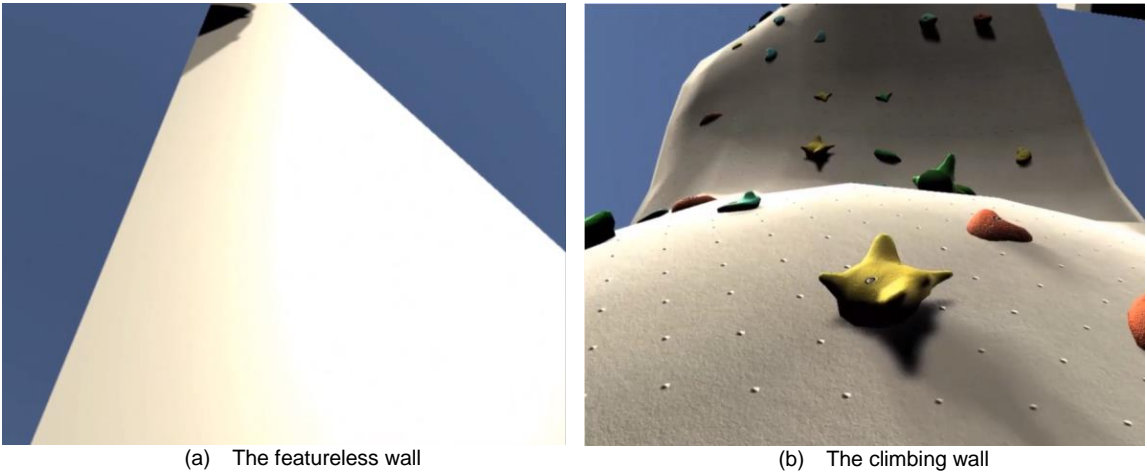

(a) The featureless wall  (b) The climbing wall

Figure 2: Participants watched a video of a person using one or two motion controllers to climb a featureless wall (a) or a climbing wall (b). The user could grab onto any part of the featureless wall but had to grab onto specific holds for the climbing wall.

The prototypes were developed using `VRTKv4` (Extended Reality Ltd.) in `Unity2019.4.2`. The Oculus Rift S headset and motion controllers were used to test and videotape the prototypes. The `grip` action was used to interact with the wall with the controlled right hand and the `trigger` action was used to trigger the action of the virtual left hand.

### 5.2.1 Prototyping: Infer Virtual Hand

The input technique where the movement of the virtual hand is inferred from the movement of the controlled hand was implemented slightly differently for the two symmetric out-of-phase interaction instances. For the climbing wall with holds, the virtual hand was placed on the hold closest to the user's headset (Figure 3a). For the featureless wall, the virtual hand was placed 0.4 m higher vertically and 0.2 m away horizontally from the controlled hand. For both walls, the virtual hand was activated upon trigger press. When the user lets go of the wall with their controlled hand, the position of the user's body was moved up in a manner that mimicked pulling the body up the wall with the virtual hand.

### 5.2.2 Prototyping: Alternative Input

Although many alternative input options were available, we created a prototype where the virtual hand was controlled by the position of the headset using raycasting, as it required no additional equipment. We added a straight pointer to the headset such that the virtual hand was attached to the end of the pointer. The pointer was enabled by pressing and holding the trigger on the motion controller, and the user-controlled the placement of the virtual hand by moving the headset (Figure 3b). The user could place the virtual hand on a desired location by releasing the trigger. When the user lets go of the wall with their controlled hand in VR, the user was moved up by the same distance that they climbed with their controlled hand to simulate pulling up the body with the virtual hand.

*5.2.3 Prototyping: Mode Switch*

We implemented the mode switch prototype by enabling the two hands in VR to be controlled by the user's right hand holding the motion controller (Figure 3c). Although either controller could be used to control the two hands in VR, our prototype assumes that the user is using their right hand to hold the controller. When the climbing task starts, the user controls the right hand in VR using the motion controller. The user can grasp a desired hold with their right hand in VR by pressing the grip button on the motion controller. Next, to start controlling the virtual left hand, the user can press and hold the trigger to disassociate the movements of the motion controller from the movements of the right hand in VR. Then, the user can move the motion controller over to the position of the virtual left hand and can start controlling the position of the virtual left hand once they release the trigger. The user can place the virtual left hand at a desired hold by pressing the grip button, and then press and hold the trigger button to disassociate the movements of the motion controller from the movements of the virtual left hand. In this way, one motion controller being held by the user's right hand can control both the left and right hands in VR.

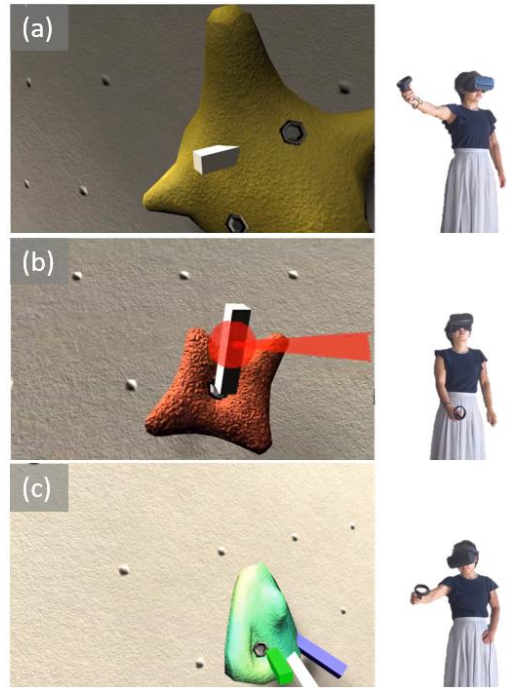

Figure 3: A screenshot of the input technique prototypes being used by a person using one motion controller in their right hand to climb up a climbing wall. (a) A prototype of inferring the virtual hand where the virtual hand appears on the closest hold upon trigger press. (b) A prototype of alternative input where the position of the virtual hand can be controlled by a red

straight ray coming out of the headset. (c) A prototype of mode switch where the position of the two hands in VR can be controlled with one motion controller sequentially.

## 5.3 Remote Video Elicitation Study Methods

The objective of the study was to assess and improve our understanding of the tradeoffs associated with having computer assistance on or off with a remote video elicitation. In addition, we wanted to highlight the need for more accessible bimanual interactions in VR and obtain feedback on our prototyped input techniques to provide a starting point for any VR application developer who may be looking for how to make their applications more accessible.

### 5.3.1 Participants

We recruited participants who enjoy gaming and have limited upper-body mobility to participate in our study. Participants were recruited through a nonprofit organization that provides support and services to people with limited motion. We screened participants using the following criteria: 1) had limited motion that affects their ability to use two motion controllers, 2) do not experience significant motion sickness when watching videos and movies, 3) were between the ages of 18 – 90, and 4) were located in the United States. Each study lasted about an hour on Microsoft Teams and participants were compensated with a $50 Amazon gift card.

Seventeen participants with various limited mobility participated in the study (Table 6; *gender:* 1 non-binary, 4 women, 12 men; *average age [mean ± standard deviation]:* 33 years ± 8.9 years). All but one participant (P10) indicated that they would experience challenges interacting with two motion controllers at the same time without accessibility adaptations. P10 used a wheelchair and discussed how they would have no issues with bimanual VR interactions as long as the range was not too large. Some participants had limited mobility in only one side of their body (P01, P02, P08, P09) whereas other participants had limited motion in both hands.

Table 6: Participant's self-reported limited mobility.

| ID | Age | Gender | Movement Constraints | Self-Reported Right Hand Movement Limitations | Self-Reported Left Hand Movement Limitations |
|---|---|---|---|---|---|
| P01 | 33 | Woman | Fetal varicella | None | Under-developed with limited range of motion and fatigue |
| P02 | 26 | Man | Cerebral palsy | None | Very limited motion |
| P03 | 30 | Man | Arthrogryposis | Some difficulty with finger dexterity | Limited motion in fingers |
| P04 | 34 | Man | C4-6 incomplete spinal cord injury | Limited motion except in shoulder and wrist, some movement in fingers | Limited motion except in shoulder and wrist, some movement in fingers |
| P05 | 18 | Man | C5 spinal cord injury | Limited wrist and finger movement | Limited wrist and finger movement |
| P06 | 61 | Man | ALS | Difficult to holds hands up against gravity | No motion |
| P07 | 33 | Man | Cerebral palsy | Limited range of motion and finger dexterity | Limited range of motion and finger dexterity, fatigue |
| P08 | 30 | Woman | Cerebral palsy | None | Limited motion, some difficulty with finger dexterity |
| P09 | 29 | Man | Stroke | Limited finger dexterity | None |
| **P10** | 25 | Man | Cerebral palsy affecting legs – uses wheelchair | None | None |
| **P11** | 39 | Non-binary | Spinal muscular atrophy | Motion limited to chest area, limited finger motion | Motion limited to chest area, limited finger motion |
| **P12** | 36 | Woman | Progressive neuromuscular disorder | Motion limited to thumb movement | Motion limited to thumb movement |
| **P13** | 33 | Woman | C4-5 spinal cord injury | Limited finger movement | No motion |
| **P14** | 33 | Man | C5 spinal cord injury | Limited finger and arm motion | Limited finger and arm motion |
| **P15** | 40 | Man | Becker muscular dystrophy | Wrist and arm fatigue | Wrist and arm fatigue |
| **P16** | 32 | Man | Duchenne's muscular dystrophy | Limited movement in hands and fingers | Limited movement in hands and fingers |
| **P17** | 31 | Man | Duchenne's muscular dystrophy | No motion | Limited movement in arms, some finger movement |

### 5.3.2 Interview Protocol

Our remote interview was constructed to elicit thoughts and opinions from users with and without VR experience (see supplementary information for sample interview questions). Due to social distancing restrictions affecting in-person user studies, we elected to collect user opinions remotely via a video elicitation study. During visual elicitation studies, a stimulus such as a photograph or a video is used to prompt participants to talk about potential scenarios in detail [McNely, 2013]. In this study, we showed participants videos of a person using different methods to climb up a wall and asked participants follow-up questions. There were two types of interaction instances (featureless wall, climbing wall) and three prototyped input techniques (infer virtual hand, alternative input, mode switch), as well as the standard dual controller input technique as a baseline for a total of eight videos created for the study. Each video was approximately one minute in length and had a side-by-side of a real-life video of a person performing the climbing interaction with the input technique and the virtual view from the person's head-mounted display.

We grouped videos corresponding to each interaction instance together and randomized both the order that the instances were presented to the participant as well as the order of the three prototyped input techniques. The video of using two controllers to interact in VR was presented at the beginning of each interaction instance to obtain a baseline. After watching each video, participants were asked how likely it was that they would use the method to perform the climbing interaction, how easy or hard the interaction would be for them to complete, and about any potential challenges they may encounter when using the method to complete the interaction. At the end of the study, participants were asked how they would adapt the one-handed method to make it more accessible to them, their interaction preferences, and whether they would consider using the single motion controller interaction method to use VR technology.

### 5.3.3 Analysis

We analyzed interview transcripts using reflexive thematic coding [Braun et al., 2006]. Two authors (MY, MM) independently read through all the interview transcripts to take notes and develop codes. Codes were discussed between the two authors until consensus was reached. One author (MY) then applied the final codes to all the transcripts, and the two authors (MY, MM) met to discuss the themes and results.

**5.4 Video Elicitation Study Results**

The main themes found in our analysis were: 1) using inaccessible gaming devices is a frustrating experience; 2) navigating tradeoffs between user effort and autonomy varies among participants; 3) additional adaptations would improve accessibility and user experience. Our results suggest that while the evaluation lens can be useful for understanding general tradeoffs between different input techniques, additional user evaluation is important for refining the developed input techniques. We highlight the need for personalizable accessibility settings for VR applications.

*5.4.1 Using Inaccessible Gaming Devices is a Frustrating Experience*

Eleven participants (P01, P03-P08, P12, P16) discussed the frustration of playing video games that are difficult or impossible to access for them. P04 mentions the challenge of playing multiplayer games on PlayStation 4: *"I found [playing Grand Theft Auto with friends] very difficult to have any competitive ability because the controls are so difficult…it's been frustrating"*. Our participants have *"…to work so much harder to [play the game] correctly"* (P08) due to their spatial and coordination difficulties and gaming often takes them *"…longer than it regularly would"* (P05) for someone who is not disabled, making gaming a frustrating experience.

With VR, using two motion controllers is a particularly challenging experience for all but one (P10) of our participants. After watching the video of a person performing the climbing motion with two motion controllers, participants listed a variety of challenges associated with using two motion controllers. The participants highlighted issues such as holding the controller with the affected arm (P01, P02, P14, P17), difficulty pressing buttons (P03, P05, P08-P09, P13-P14), fatigue (P04-P08, P13-P14), and range of motion (P03, P05-P08 P10-15, P15-P16). For example, P14 discussed how *"trying to hold both [motion controllers]* **and** *hit a button would be difficult to do"* and P01 mentioned: *"I couldn't do [bimanual interactions in VR] because I can't hold the second controller…there's no point [in playing VR games] because it would make me upset"*.

Three participants (P02, P09, P14) indicated that they would have bought or seriously considered purchasing a VR headset if the technology was accessible for their abilities. P05 and P06 discussed how they own a VR headset but do not use it due to inaccessibility and P03 previously owned a VR headset but gave it away because they could not press the motion controller buttons.

Accessibility challenges were not only related to the VR device itself. For example, P09 discussed how accessibility can be tied to specific games: *"when I go in to play games at my brother's [house] I'm always like, you know me, you know what I can and can't do, what games would you suggest that I play that I can do one-handed?"* P07 also mentioned: *"there has to be a way to communicate that these games are for everyone"* and not just for people who can use two motion controllers.

*5.4.2 Navigating Tradeoffs Between User Effort and Autonomy Vary Among Participants*

All but one (P10) participant thought that the single-handed motion controller interactions would make VR more accessible to them. When asked which of the three prototypes they preferred, eight participants (P03, P05-P07, P11-P13, P15) preferred the input technique where the computer inferred the movement of the virtual hand, nine preferred the alternative input (P01-P02, P04, P08-P10, P15-P17), and two preferred the mode switch (P12, P14).

Participants preferred the option to infer the movement of the virtual hand because it appeared easier, less frustrating, seemed intuitive, and *"takes only half as much energy"* (P13). P05 felt that inferring the movement of the virtual hand provided *"the smoothest experience…I want to do it as normally as possible and have very little accessible options that would take me longer than it regularly would so I think this would make it more seamless"*. However, participants cited decreased autonomy as a drawback, including not being able to choose the position of the virtual hand, potential frustration with the game not knowing what the user wants to do, taking the fun out of the game, and the method feeling like cheating. However, P01 mentions that they *"would rather participate partially than not at all"* and would use this method if that was the only accessibility option available. P01 and P14 suggested providing the option of toggling between a couple of different options as a work-around to the issue of the game not correctly inferring the user's ideal position of the virtual hand.

Nine participants preferred the alternative input prototype of using the headset to control the position of the virtual hand. Some of their reasons included that the method provides less fatigue and strain compared to using two motion controllers, it provides the smoothest user experience, and it enables users to control movements enjoyably. P08 mentioned *"that is super dope, it seems easier…[the game is] not just doing it for me, I still have to mark the position [of the virtual hand] with the laser…but it seems like a lot less coordination on my part…[I'm] still getting the same experience… it's just a different set of inputs for [the same interaction]"*. Other participants, however, said that alternative input was not preferred because it is confusing, takes them out of the immersion of performing the movement, and could cause neck fatigue and motion sickness. P11, P13, P16, and P17 mentioned challenges with moving their head and neck but said that they would want to use this method if they could use eye gaze instead of head gaze to control the virtual hand.

P12 and P14 preferred the mode switch because it enables them to use VR technology and *"it gives you more control"* (P12). However, participants were concerned that the method was too complicated, it would take too much time to complete different movements, it would take them out of the immersion of VR, and it would cause too much fatigue because they are using one hand to control the motion of two hands. Some participants suggested that it *"doesn't add much to the interaction to know that I can control both hands"* (P15) and *"loses the purpose"* (P05) of the interaction because of the perceived difficulty of the method. This sentiment was particularly strong for the featureless wall where hand placement does not significantly affect gameplay.

Participants preferred different adaptations based on their preferred level of control over the virtual hand. Participants who were comfortable with the *"developer holding my hand and guiding me through the experience as much as possible"* (P15) tended to prefer more automated options, like the option to infer the movement of the virtual hand. Other participants said that it would be dependent on the movement required, and how much having more control over the virtual hand would slow down the game speed, and that *"there are definitely situations where I think [the VR game] could be more automated, it would be cool if there was an option to turn that on and off"* (P14). Although most participants had one or two prototypes that they preferred over the others, participants wanted as many accessibility settings as possible so that they could customize their inputs based on how much autonomy they want over their avatar and their pain and fatigue levels.

*5.4.3 Additional Adaptations Would Improve Accessibility and User Experience*

In addition to the accessibility options presented by the three prototypes, participants suggested additional adaptations that would make VR interactions more accessible to them. P06, P11-P13, P15-P17 indicated that they would experience significant challenges with range of motion with either hand, as their movements are limited to moving their fingers in one or two hands. These participants wanted to use the inputs that they normally use to control a computer (e.g., mouse, touchpad, or joystick) or other inputs (e.g., foot, eye gaze) for the controlled hand instead of a motion controller and use the proposed alternative methods to control the virtual hand. P06 discussed: *"I would really like to see [VR developers] try to take advantage of moving your feet because I can lift and move my feet much better than my arms"*. P16 also discussed how having alternative methods to control one of the hands in VR is beneficial *"because it lowers the number of buttons and switches you have to use"* when using alternative inputs like switches and quad sticks.

Remapping buttons was also a highly requested feature from 8 participants. P01, P03, P05, P09, P11-P14 suggested remapping buttons from the motion controller to something that can be pressed by their other hand, with a foot pedal, or with a shoulder button due to difficulty pressing buttons with their dominant hand. Participants also suggested toggling buttons so that they do not have to hold and press a button to activate or inactivate gripping and triggering.

For participants with limited range of motion, *"making the [movement] sensitivity something that could be adjustable"* (P08) was suggested to *"have the movement needed scaled back considerably"* (P12). Additionally, P12 mentioned, *"I would love to be able to just go in and interact with things like I would in normal life"* using adaptations learned in everyday life to make everyday interactions more accessible. Lastly, participants with limited neck movement suggested using eye gaze instead of head gaze as an alternative input to control the virtual hand (P03, P06, P07, P11, P16, P17). P16, whose limited mobility prevents them from moving their head discussed: *"to use that range of motion with my head, that would be very difficult…the eye gaze would solve a lot of the moving the head around…[if I used eye gaze] I could just look at it and click one button and go"*.

## 6 DISCUSSION

We conducted a user study to assess and improve our understanding of the tradeoffs associated with having computer assistance on or off when performing bimanual tasks, and to highlight the need for more accessible bimanual interactions in VR. Our user study reinforces how inaccessible VR is for people with a wide variety of mobility limitations and demonstrates how input techniques can be designed to make VR motion controllers more accessible. Prior work has found that interacting with two motion controllers simultaneously is a difficult or impossible task for many people with limited mobility and deters potential users from interacting with VR technology [Mott, 2020]. However, making VR bimanual interactions accessible appears to be a daunting and endless task for VR application developers. Our Two-In-One design space provides a framework for considering how groups of bimanual interactions can be made accessible for people with limited mobility. The creation lens enables the development of input techniques that may leverage computer assistance and the evaluation lens provides some insight into the tradeoffs that users experience when using different input techniques. In this section, we discuss observations from our user study, insights from our experience constructing our Two-in-One design space, and we contextualize these observations and insights with findings from prior work.

### 6.1 Customizable Input Techniques is Important for Accessibility

As we expected, dual motion controllers were inaccessible to all our participants with upper-body limited mobility. Regardless of whether they could or could not accessibly use the input techniques as they were currently prototyped, all but one of our participants felt that enabling bimanual input with one motion controller would improve VR accessibility.

We initially conceived the input techniques under the assumption that the potential user would be able to fully control one motion controller and have limited or no mobility in the other, such as people who had a stroke affecting one side of the body, people who had limb loss, or people who needed one hand to manipulate their wheelchair movement. Our participants who fit this mobility category felt they could access the prototyped input technique without additional adaptations. Although our participants generally wanted autonomy in gameplay and less reliance on computer assistance, they discussed how that would have to be balanced against frustration at not being able to play the game at a reasonable pace, especially if it was a game that required gameplay at a certain speed. Many of these participants had one input technique they preferred but wanted access and the ability to switch between all three prototyped input techniques depending on the VR application and the participants' fluctuating mobility limitations. For example, some participants suggested if a movement is not essential to the gameplay, such as pouring medicine into a cup to regain health during an action game, automating such interactions could help ensure that users are focused on the experience rather than the movements. However, for movements that are essential to the strategy of the game or application, such as pouring drinks in a certain order for a restaurant simulator, it may be beneficial to enable full control over the movement. Two-In-One could help guide which techniques to choose over others based on the goals of the interaction.

In addition to making VR motion controllers more accessible to people with limited motion in one hand—by lowering the number of handheld input devices needed from two to one—we also made bimanual interactions in VR more accessible for people who experience fatigue, have difficulty pressing buttons, and whose hand movements are limited to the use of specific input devices (e.g., mouse or joystick) with additional adaptations. For people who experience fatigue, interacting in VR with one motion controller at a time increases the time they can interact in VR. For people with difficulty pressing buttons, they can use one hand to control the motion of the controller and the other to press buttons to move and interact in VR. For people whose movements are limited to their fingers, decreasing the degrees of freedom they need to control from two hands in VR to one hand decreases the number of alternative input devices needed to remap the motion controller movements to other input devices like mice or joysticks. Decreasing the number of input devices required to interact in VR is useful for expanding VR accessibility to people with diverse motor abilities and presents new opportunities for developing pluggable VR inputs to improve VR motion controller accessibility.

### 6.2 Two-In-One as an Iterative Design Process

In this work, we focused on one iteration of applying the creation and evaluation lens and then demonstrating the prototypes to potential users. However, we envision the application of the design space to be more of an iterative process, applying the user feedback to develop more accessible input techniques. Additionally, the Two-In-One design process and potential input techniques suggested here can be adapted to developer or user preference or experiential goals of the gameplay.

Many design choices can be made that impact usability of the techniques for different abilities. For example, in our implementation of mode switch, the user triggered the switch from controlling one virtual hand to another by pressing and holding the trigger. There are other ways in which this input technique could have been implemented – for example, the mode switch could be triggered automatically depending on which side of the user's body the controller is located. If the user's controller is located on the right side of the body, it could be inferred that the user is trying to control the right hand in VR, and vice versa. Other design choices like whether the virtual hand snaps to the position of the controller may or may not impact gameplay depending on the abilities of the user. For example, if a user has difficulty with range of motion, it may be easier for them to control the relative position, rather than the absolute position of the virtual hand so they do not have to cross over their body to control the virtual hand. However, for a user who does not have limited range of motion, controlling the absolute position of the virtual hand may be preferable to improve immersion and usability. In this way, design choices can be made iteratively with user input being incorporated as the input techniques are developed and refined. Providing users with a variety of options that can be personalized to fit their abilities is the best way to circumvent creating inaccessible games and applications.

### 6.3 Beyond VR Motion Controller Accessibility for People with Disabilities

Although this study focused on improving VR motion controller accessibility for people with long-term limited mobility, the Two-In-One design space could be extended for people with related short-term limited mobility (e.g., breaking an arm) or situational impairments (e.g., holding a child) [Sears et al., 2003]. Two-In-One also provides potential avenues for improving rehabilitation through eliciting the self-avatar follower effect for the virtual hand, where an individual is influenced by an avatar's motions [Gonzalez-Franco et al., 2020]. For example, asking participants with nerve damage and phantom upper limb pain to view their phantom limb moving in VR resulted in significantly reduced pain [Osumi et al., 2016]. Additionally, the presented work could further extend the abilities of users using two hands to interact in VR to further control additional degrees of freedom. For example, Two-In-One principles can be applied towards creating a VR application for users without limited mobility to virtually control a multi-armed robot or animal avatar like an octopus [Won et al., 2015]. Although the work presented here was inspired by a need for a specific group of users, Two-In-One could also expand capabilities for VR users without limited mobility that are not yet possible with current VR technology.

Although this paper focused on improving the accessibility of dual motion controllers for VR, this work is relevant for other extended reality technologies and assistive technology in the real world as well. Augmented reality devices such as the HoloLens rely heavily on two-handed manipulations and are already starting to be deployed in work environments. Lastly, our work could also extend back to accessibility challenges in the real world as well. Bimanual interactions are a challenge for many people with limited mobility in the real world, and the performance of assistive technology like prosthetics and orthotics could be improved by incorporating the object-based input techniques suggested in the Two-In-One design space. Such extensions provide exciting opportunities to improve accessibility in many areas of life, not just during VR use.

### 6.4 Computer Assistance Provides a Unique Advantage for VR Accessibility

Our demonstration that all bimanual interactions in VR can be categorized into one of four interaction types may be a surprising one – and one that enables unique opportunities for enabling VR accessibility via computer assistance. Because of the limited number of interaction types possible, VR application designers and developers can create input techniques that take advantage of the digital nature of VR. There is an unlimited number of possibilities to develop computer-assisted input techniques to address VR motion controller accessibility challenges beyond the ones that we considered here. For example, developing machine learning models and data-based techniques to infer the movement of the virtual hand could make the computer assistance feel much more realistic than the heuristic-based techniques we developed in this paper. As processing power gets cheaper and more available, the ability to make VR applications more accessible will continue to get easier, and it is important to consider how we can take advantage of such technology improvements as they occur.

## 6.5 Limitations

Our main limitation was that our participants could not experience the prototypes first-hand and inferred their preferences from the video elicitation. Because immersion is such an important aspect of VR, we expect that participants' opinions might change if they were to experience the prototypes firsthand. Additionally, only a few of our participants had first-hand experience using VR motion controllers, so many had to infer the accessibility of the motion controllers from the videos. Our user study predominantly focused on men's perspectives– we only had four people who identified as women and one person who identified as non-binary in our study – and it would have been beneficial to have broader gender perspectives. We created a limited number of prototypes with two instances of climbing. Although the limited interaction instances helped in identifying participant preferences for a particular interaction (symmetric out-of-phase), we did not identify participant preferences for other interactions, such as symmetric in-phase, asymmetric coordinated, and asymmetric uncoordinated. Lastly, because we surveyed the most popular VR applications overall rather than popular applications in each category, most of the applications included in the survey were action games. It would have been beneficial to examine bimanual interactions present in other types of applications, such as meditation or art.

## 7 CONCLUSION

We presented Two-in-One, a design space to enable bimanual interactions in VR with unimanual input. We used the design space to classify bimanual interactions in popular VR games into one of four categories: symmetric in-phase, symmetric out-of-phase, asymmetric coordinated, and asymmetric uncoordinated. We then demonstrated that each interaction type has properties that can be leveraged to develop input techniques that enable bimanual VR interactions given one motion controller input with or without computer assistance. We prototyped three input techniques and assessed user preferences with a video elicitation study for two symmetric out-of-phase interactions. People with limited mobility have different preferences for how they want to interact in VR and their preferences are dependent on the VR application and daily changes in needs and abilities. We present additional adaptations that can further enhance the accessibility of VR for people with limited mobility. Two-in-One provides a blueprint for how bimanual interactions in VR can be enabled with one motion controller and our user study highlights the need and provides a starting place to develop a diverse set of accessibility tools to make VR motion controllers accessible to a broad range of abilities.

## ACKNOWLEDGMENTS

Thank you to all our participants who provided valuable input to our study, and to the AbleGamers Foundation for their help in participant recruitment.

**REFERENCES**

2020. Specialeffect: The Gamers' Charity.

2MW 2002. WalkinVRDriver.

Ahuja, K., Ofek, E., Gonzalez-Franco, M., Holz, C. and Wilson, A.D. 2021. Coolmoves: User Motion Accentuation in Virtual Reality. *Proceedings Of The ACM On Interactive, Mobile, Wearable And Ubiquitous Technologies 5*, 1-23.

Annett, M. 1970. A Classification of Hand Preference by Association Analysis. *British Journal of Psychology 61*, 303-321.

Braun, V. and Clarke, V. 2006. Using Thematic Analysis in Psychology. *Qualitative Research in Psychology 3*, 77-101.

Brudy, F., Holz, C., R Dle, R., Wu, C.J., Houben, S., Klokmose, C.N. and Marquardt, N. 2019. Cross-Device Taxonomy: Survey, Opportunities and Challenges of Interactions Spanning Across Multiple Devices. In *Proceedings Of The 2019 CHI Conference on Human Factors in Computing Systems*, 1-28.

Buxton, W. and Myers, B. 1986. A Study in Two-Handed Input. *ACM SIGCHI Bulletin 17*, 321-326.

Cai, F. and Rijke, M.D. 2016. A Survey of Query Auto Completion in Information Retrieval. *Foundations and Trends® in Information Retrieval 10*, 273-363.

Card, S.K., Mackinlay, J.D. and Robertson, G.G. 1990. The Design Space of Input Devices. In *Proceedings Of The SIGCHI Conference on Human Factors in Computing Systems*, 117-124.

Card, S.K., Mackinlay, J.D. and Robertson, G.G. 1991. A Morphological Analysis Of The Design Space Of Input Devices. *Acm Transactions On Information Systems (Tois) 9*, 99-122.

Charity, T.A. 2020. Able Gamers.

David-John, B., Peacock, C., Zhang, T., Murdison, T.S., Benko, H. and Jonker, T.R. 2021. Towards Gaze-Based Prediction of the Intent to Interact in Virtual Reality. In *ACM Symposium On Eye Tracking Research and Applications*, 1-7.

Gerling, K., Dickinson, P., Hicks, K., Mason, L., Simeone, A.L. and Spiel, K. 2020. Virtual Reality Games for People Using Wheelchairs. In *Proceedings Of The 2020 CHI Conference on Human Factors in Computing Systems*, 1-11.

Gerling, K. and Spiel, K. 2021. A Critical Examination of Virtual Reality Technology in the Context of the Minority Body. In *Proceedings Of The 2021 CHI Conference on Human Factors in Computing Systems*, 1-14.

Gonzalez-Franco, M., Cohn B., Ofek E., Burin D. and Maselli A. 2020. The Self-Avatar Follower Effect in Virtual Reality. In *2020 IEEE Conference on Virtual Reality and 3D User Interfaces (VR),* 18-25.

Guiard, Y. 1987. Asymmetric Division Of Labor In Human Skilled Bimanual Action: The Kinematic Chain As A Model. *Journal of Motor Behavior 19*, 486-517.

2020. Game Accessibility Guidelines.

Haaland, K.Y., Mutha, P.K., Rinehart, J.K., Daniels, M., Cushnyr, B. and Adair, J.C. 2012. Relationship Between Arm Usage and Instrumental Activities of Daily Living After Unilateral Stroke. *Archives of Physical Medicine and Rehabilitation 93*, 1957-1962.

Hinckley, K., Czerwinski, M. and Sinclair, M. 1998. Interaction and Modeling Techniques for Desktop Two-Handed Input. In *Proceedings Of The 11th Annual ACM Symposium on User Interface Software and Technology*, 49-58.

Hinckley, K., Pausch, R., Proffitt, D., Patten, J. and Kassell, N. 1997. Cooperative Bimanual Action. In *Proceedings Of The ACM SIGCHI Conference on Human Factors in Computing Systems*, 27-34.

Hinckley, K., Pierce, J., Sinclair, M. and Horvitz, E. 2000. Sensing Techniques For Mobile Interaction. In *Proceedings Of The 13th Annual ACM Symposium on User Interface Software and Technology*, 91-100.

Hirzle, T., Gugenheimer, J., Geiselhart, F., Bulling, A. and Rukzio, E. 2019. A Design Space for Gaze Interaction on Head-Mounted Displays. In *Proceedings Of The 2019 CHI Conference On Human Factors in Computing Systems*, 1-12.

Horvitz, E. 1999. Principles of Mixed-Initiative User Interfaces. In *Proceedings of the SIGCHI Conference on Human Factors in Computing Systems*, 159-166.
26